\documentclass{ws-jai}
\usepackage[flushleft]{threeparttable}
\usepackage{graphicx}
\usepackage{soul,xcolor}
\usepackage[tableposition=top]{caption}
\usepackage{subcaption}
\usepackage{url}
\usepackage[colorlinks=true,allcolors=blue]{hyperref}
\usepackage{natbib}
\usepackage{floatrow}

\def\albatros{ALBATROS}
\def\mars{MARS}

\setstcolor{red}

\newfloatcommand{capbtabbox}{table}[][\FBwidth]

\begin{document}

\catchline{}{}{}{}{} 

\markboth{T.~Dyson et al.}{Radio-Frequency Interference at the McGill Arctic Research Station}

\title{Radio-Frequency Interference at the McGill Arctic Research Station}

\author{T.~Dyson$^{1,\dagger}$, H.~C.~Chiang$^{1,2}$, E.~Egan$^{1}$,
  N.~Ghazi$^{2}$, T.~M\'enard$^{1}$, R.~A.~Monsalve$^{1,3,4}$, T.~Moso$^{2}$, J.~Peterson${^5}$,
  J.~L.~Sievers$^{1,6}$, S.~Tartakovsky$^{1}$}

\address{
$^{1}$Department of Physics, McGill University, Montr\'eal, Qu\'ebec H3A 2T8, Canada\\
$^{2}$School of Mathematics, Statistics, and Computer Science, University of KwaZulu-Natal, Durban 4000, South Africa\\
$^{3}$School of Earth and Space Exploration, Arizona State University, Arizona 85287, USA\\
$^{4}$Facultad de Ingenier\'ia, Universidad Cat\'olica de la Sant\'isima Concepci\'on, Alonso de Ribera 2850, Concepci\'on, Chile\\
$^{5}$Department of Physics, Carnegie Mellon University, Pittsburgh, Pennsylvania 15213, USA\\
$^{6}$School of Chemistry and Physics, University of KwaZulu-Natal, Durban 4000, South Africa
}

\maketitle

\corres{$^{\dagger}$Corresponding author: taj.dyson@mail.mcgill.ca}

\begin{history}
\received{(to be inserted by publisher)};
\revised{(to be inserted by publisher)};
\accepted{(to be inserted by publisher)};
\end{history}

\begin{abstract}
The frequencies of interest for redshifted 21~cm observations are heavily affected by terrestrial radio-frequency interference (RFI). We identify the McGill Arctic Research Station (\mars) as a new RFI-quiet site and report its RFI occupancy using 122~hours of data taken with a prototype antenna station developed for the Array of Long-Baseline Antennas for Taking Radio Observations from the Sub-Antarctic.  
Using an RFI flagging process tailored to the \mars\ data, we find an overall RFI occupancy of 1.8\% averaged over 20--125~MHz. In particular, the FM broadcast band (88--108~MHz) is found to have an RFI occupancy of at most 1.6\%. 
The data were taken during the Arctic summer, when degraded ionospheric conditions and an active research base contributed to increased RFI. The results quoted here therefore represent the maximum-level RFI environment at \mars.
\end{abstract}

\keywords{radio astronomy; site testing}

\section{Introduction}

One of the greatest challenges facing contemporary radio astronomy
experiments is terrestrial radio-frequency interference (RFI), which
has steadily worsened over time as the globe has been populated with
an increasing number of transmitters and other radiating sources.
Radio astronomy experiments are often forced to operate from remote
locations, where the RFI background is minimized, but the remoteness
requirement is directly at odds with the simultaneous need for
accessibility and logistical infrastructure.  The need for RFI-quiet
locations is especially important for cosmic dawn experiments
measuring globally averaged 21-cm emission of neutral
hydrogen~\citep{2020arXiv200510669D, 2019ApJ...883..126N,
  2019JAI.....850004P, 2018Natur.555...67B, 2018ExA....45..269S}.
Because these experiments measure total power, the requirements on
background RFI levels are far more stringent than the typical
requirements for interferometric experiments, which benefit from
cross-correlation.  The problem is further compounded by the observing
frequency range for cosmic dawn ($\sim 30$--200~MHz), which
encompasses the FM broadcast band.

We have identified the McGill Arctic Research Station (\mars)\footnote{\url{https://www.mcgill.ca/mars/}}
as a new location with an exceptionally quiet RFI environment that can serve as
an observing site for future low-frequency radio astronomy
experiments.  This paper presents spectral measurements below 125~MHz
that were taken from \mars\ during July~2019, the methodology for
identifying RFI in the data, and an assessment of the RFI
occupancy.

\section{Instrument}\label{sec:instrument}

\begin{figure}
  \begin{center}
    \includegraphics[width=0.8\linewidth]{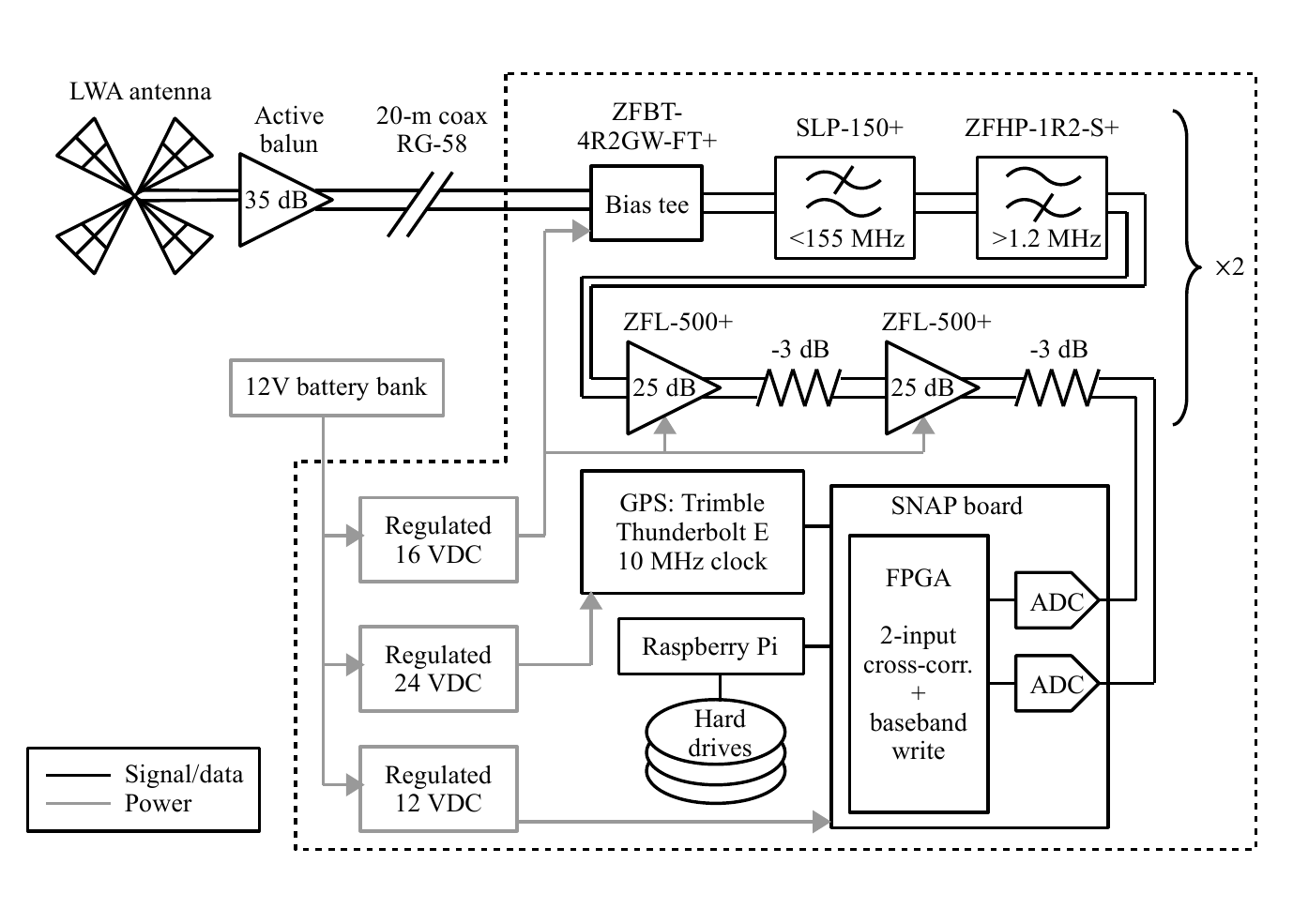}
    \caption{Block diagram of the \albatros\ pathfinder antenna
      installed at \mars.  A dual-polarization LWA antenna, equipped
      with a front-end active balun, connects via 20-m coaxial cables
      to the back-end readout electronics, housed in a Faraday cage
      denoted by the dashed lines.  Each of the two antenna signals is
      passed to a second-stage electronics chain consisting of filters
      and further amplification.  The signals are digitized at
      250~Msamp/s by a SNAP board, which computes channelized baseband
      data and spectra.  The system is powered by 12~V batteries that
      are manually recharged.}
    \label{fig:albatros}
  \end{center}
\end{figure}

The primary data presented in this paper were recorded by a single
pathfinder antenna station that was developed for the Array of Long
Baseline Antennas for Taking Radio Observations from the
Sub-Antarctic~\citep[\albatros;][]{2020arXiv200812208C}.
Figure~\ref{fig:albatros} shows a schematic diagram of the
\albatros\ station.  Incoming signals are received by a
dual-polarization Long Wavelength Array (LWA) dipole
antenna~\citep{1552313}, outfitted with an active-balun front-end
electronics (FEE) module that provides $\sim37$~dB
gain~\citep{2012PASP..124.1090H}.  The FEE is powered with 16~V, which
is passed through the coaxial cable via bias tees.  The back-end RF
electronics consist of high- and low-pass filters from Minicircuits that band-limit the
signal to 1.2--155~MHz, and amplifiers and attenuators that together
provide an additional $\sim44$~dB gain.  A Smart Network ADC
Processor~\citep[SNAP;][]{2016JAI.....541001H} board digitizes the RF
signals at 250~Msamp/s, and the ADCs are locked to a 10-MHz reference
produced by a Trimble Thunderbolt~E GPS-disciplined clock module.  The
SNAP FPGA computes auto- and cross-spectra of the two inputs over the
full 0--125~MHz frequency range, with 2048 frequency channels and
accumulation over few-second intervals.  (The FPGA also computes
channelized baseband data for each polarization over tunable frequency
windows within the 0--125~MHz operating range, but these data are not
used in the analysis presented here.)  The low-pass filter cutoff in
the back-end is intentionally set higher than the Nyquist frequency to
alias in the 137--138~MHz downlink signal from the ORBCOMM satellite
constellation.  Thus, any RFI observed above 95~MHz may have been aliased
from 125--155~MHz.  A Raspberry Pi 3B+ single board computer controls
the SNAP board and receives the auto- and cross-spectra via GPIO
connections, and the spectra are saved to an on-board SD card.
The back-end electronics (within the dashed box in
Figure~\ref{fig:albatros}) are portable, but the LWA antenna and
front-end are not. At the sites surrounding
\mars\ (Figure~\ref{fig:map}), RFI measurements were taken with a
LoWavz LW-10K60M antenna outfitted with its impedence matcher, a
Mini-Circuits ZFL-500LN+ amplifier providing $\sim24$~dB gain, and a
bias tee to block the DC voltage supplied by the back-end electronics.
While the LoWavz antenna is not sufficiently sensitive for
cosmological observations, especially above 60~MHz, it suffices for an
RFI survey.

\section{Observations}

\begin{figure}
\begin{floatrow}
\capbtabbox{
  \begin{tabular}{@{}ccc@{}} \toprule
  Site Label & Latitude & Longitude \\ 
  & ($^\circ$ N) & ($^\circ$ W) \\ \colrule
  MARS & 79.415 & 90.749 \\
  1 & 79.402 & 91.209 \\
  2 & 79.370 & 90.953 \\
  3 & 79.343 & 90.601 \\
  4 & 79.425 & 90.654 \\
  5 & 79.355 & 90.726 \\
  6 & 79.457 & 90.801\\ \botrule
  \end{tabular}
}{
  \caption{Locations of RFI survey sites}\label{tab:sites}
}
\ffigbox[\Xhsize][]{
  \includegraphics[scale=0.45]{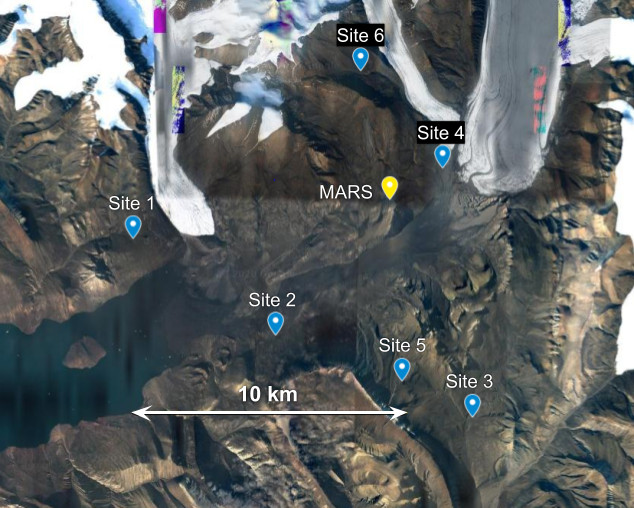}
}{
  \caption{Google Earth satellite map of Expedition Fjord with the main \mars\ base and survey sites marked. Survey sites are numbered in the order visited.}\label{fig:map}
}
\end{floatrow}
\end{figure}

\mars\ is a small research base located at $79^\circ 26'$~N, $90^\circ
46'$~W, approximately 40~km inland at the head of Expedition Fjord on
Axel Heiberg Island, Nunavut. \mars\ is accessible during
April--August via small chartered aircraft from Resolute Bay (540~km south). The closest radio transmitters are Eureka (120~km
northeast), Grise Fjord (360 km~south),  Resolute Bay, and Alert (590~km northeast), suggesting
that \mars\ and its surroundings should have low levels of
RFI. Furthermore, \mars\ is attractive for low-frequency radio
astronomy because favourable ionospheric conditions are expected at
high latitudes, especially during the Arctic winter when the polar
ionosphere is only weakly ionized in the prolonged absence of solar
radiation.  Simulations using the International Reference Ionosphere 
model~\citep{2018AdRS...16....1B} suggest that the summer and winter
plasma cutoff frequencies differ by about a factor of two at \mars. 
Additionally, measurements taken during the Arctic winter
are free of solar RFI during the continuous night. Thus, due to its
isolation from radio transmitters, its polar latitude, and its
reasonable accessibility, \mars\ is a promising candidate location for
ground-based low-frequency radio astronomy and is being investigated
as a potential site for the \albatros\ and
MIST\footnote{\url{http://www.physics.mcgill.ca/mist}} experiments.
To assess the RFI environment at \mars, observations were taken using
the instrumentation described in \S\ref{sec:instrument} at the main
\mars\ base and at several surrounding sites shown in
Figure~\ref{fig:map}.

\subsection{\mars\ base}

The RFI measurements discussed in this paper are derived
from data taken at the \mars\ base during July~9--21, 2019.  A total
of 122 and 106 hours of data were recorded from the east-west and
north-south antenna polarizations, respectively.
During the summer period, the sun remains above the horizon continually, raising the ionospheric plasma cutoff frequency in comparison to winter conditions. Furthermore, when the base is actively used by researchers, power is supplied primarily by solar panels and a power inverter, which generate a large amount of RFI. Data were therefore recorded during overnight periods while the \mars\ power system was turned off to eliminate the largest source of local RFI.  Some weaker sources of local RFI (e.g., personal electronics) remained present, particularly as the base population increased during the summer period.  Because of slightly elevated local RFI and degraded ionospheric conditions, the data presented here likely represents the worst-case RFI environment at \mars.

\subsection{Survey sites}\label{sec:survey}

To investigate potential local variation in the RFI environment, data were also recorded at several sites within 10~km of \mars, as shown in Figure~\ref{fig:map} and listed in Table~\ref{tab:sites}. These sites are candidate locations for future \albatros\ antenna installations because they have suitable terrain and are accessible by foot.
All data at these sites were taken while the \mars\ power inverter and electronics were on; however, the RFI emitted at the base is not visible at few-km separation distances. The observing period at sites 1--5 was $\sim20$ minutes each, and data were recorded at site~6 for $\sim2$~hours.

\begin{figure}
\begin{center}
\includegraphics[scale=0.7]{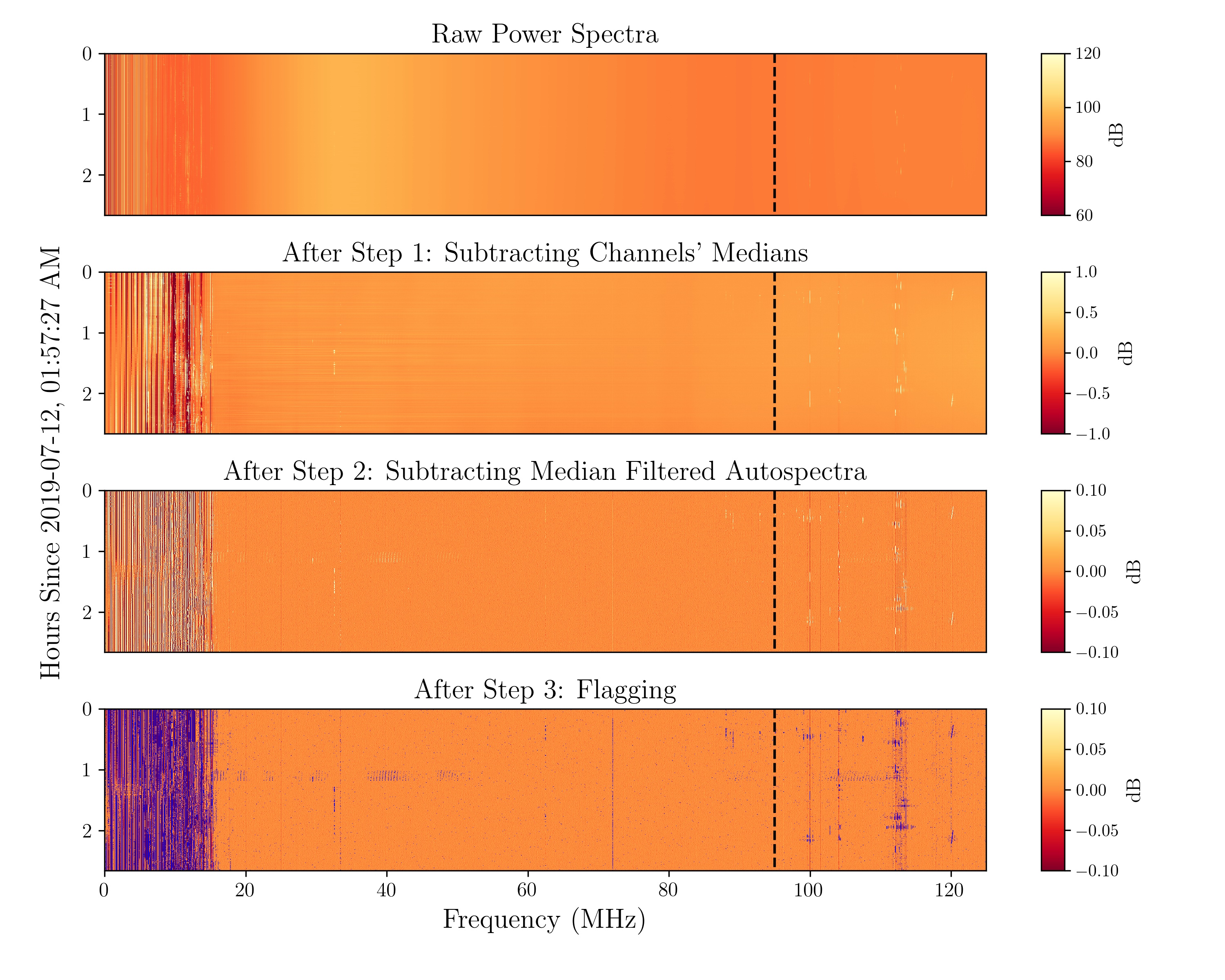} 
\end{center}
\caption{This figure shows a segment of east-west polarized autospectra recorded at \mars\ on the night of July 11\textsuperscript{th} that is representative of the entire dataset. The top panel shows the raw power spectra, while the other panels show the data after each step of the flagging process (see \S\ref{sec:method}). Flagged points are marked in blue in the final panel. The dashed line at 95~MHz indicates that any signal at higher frequency could have been aliased from 125--155~MHz (see \S\ref{sec:instrument}). Note that while only 3~hours of data are shown, the channels' medians were taken over the entire night.}
\label{fig:steps}
\end{figure}

\section{RFI identification methodology}\label{sec:method}

The \mars\ RFI environment was characterized by analyzing autospectra
with the processing steps described below, which focus on identifying
narrow-band features.  Each of the two antenna polarizations for each
night of data, comprising $\sim8$~hours of obervation, is treated
independently.  Since RFI can be many orders of magnitude brighter
than Galactic emission, the analysis procedure is performed on the
autospectra on an arbitrary decibel scale.

\begin{arabiclist}[(3)]
\item The median of each frequency channel across the duration of the observation is subtracted from that channel. This step removes most of the large-scale Galactic spectral structure, flattening the signal in preparation for the median filter applied in the next step. The first and second panels of Figure~\ref{fig:steps} show the data before and after this step, respectively.

\item Each autospectrum is passed through a median filter with a window of 300~kHz (5~channels), which roughly corresponds to the width of the narrowest non-RFI features, which are  caused by cable reflections.
Each median-filtered autospectrum is then subtracted from each original autospectrum to
remove any signal that varies arbitrarily quickly in time but only slowly in frequency,
such as the Galaxy's diurnal variation.
The data after subtracting the median filtered autospectra are shown in the third panel of Figure~\ref{fig:steps}.

\item All points more than five median absolute deviations (MADs) above the median taken over the whole observation are flagged as RFI. Flagged data are shown in blue in the bottom panel of Figure~\ref{fig:steps}.
\end{arabiclist}

After the data are flagged, the RFI occupancy is calculated as the ratio of
flagged samples to total samples for each frequency channel.
Statistical noise in the data contribute a small false flagging rate.
For Gaussian noise, 1~MAD $= 0.67 \sigma$, and the flagging
threshold of 5~MADs therefore corresponds to 3.35$\sigma$. The expected false
flagging rate, corresponding to the percentage of noise $>3.35\sigma$,
is thus 0.04\% at a minimum (variations in noise level during an observation due to 
changing Galactic brightness could cause higher false flagging rates). 

The above flagging procedure has two main limitations.  First, the
flagging breaks down for RFI-saturated frequency channels, where the
true RFI occupancy is above 50\%.  During the Arctic summer, shortwave
and HF radio reflected from the ionosphere cause this saturation at
frequencies below roughly 20~MHz.  Although RFI occupancy is
calculated over the full frequency range, the values below 20~MHz are
likely underestimated and are omitted when computing the average
levels.  Second, the flagging procedure is insensitive to
low-intensity broadband RFI features, which cannot be distinguished
from broadband non-RFI signal variation (such as gain fluctuations).
The removal of these types of signals can be seen in
Figure~\ref{fig:steps}, where faint horizontal striping features are
subtracted between the second and third panels.  Because the flagging
procedure can potentially remove bright broadband RFI as well, we
searched for these high-amplitude events separately by examining the
distribution of root mean square (RMS) values of each median filtered
autospectrum (computed as part of step~2) for frequencies above 20~MHz.
No outliers were detected, and the median and maximum RMS values are 0.1 dB and 0.3 dB respectively over all
nights at \mars, thus confirming the absence of bright broadband RFI.

\begin{figure}
\begin{center}
	\includegraphics[scale=0.8]{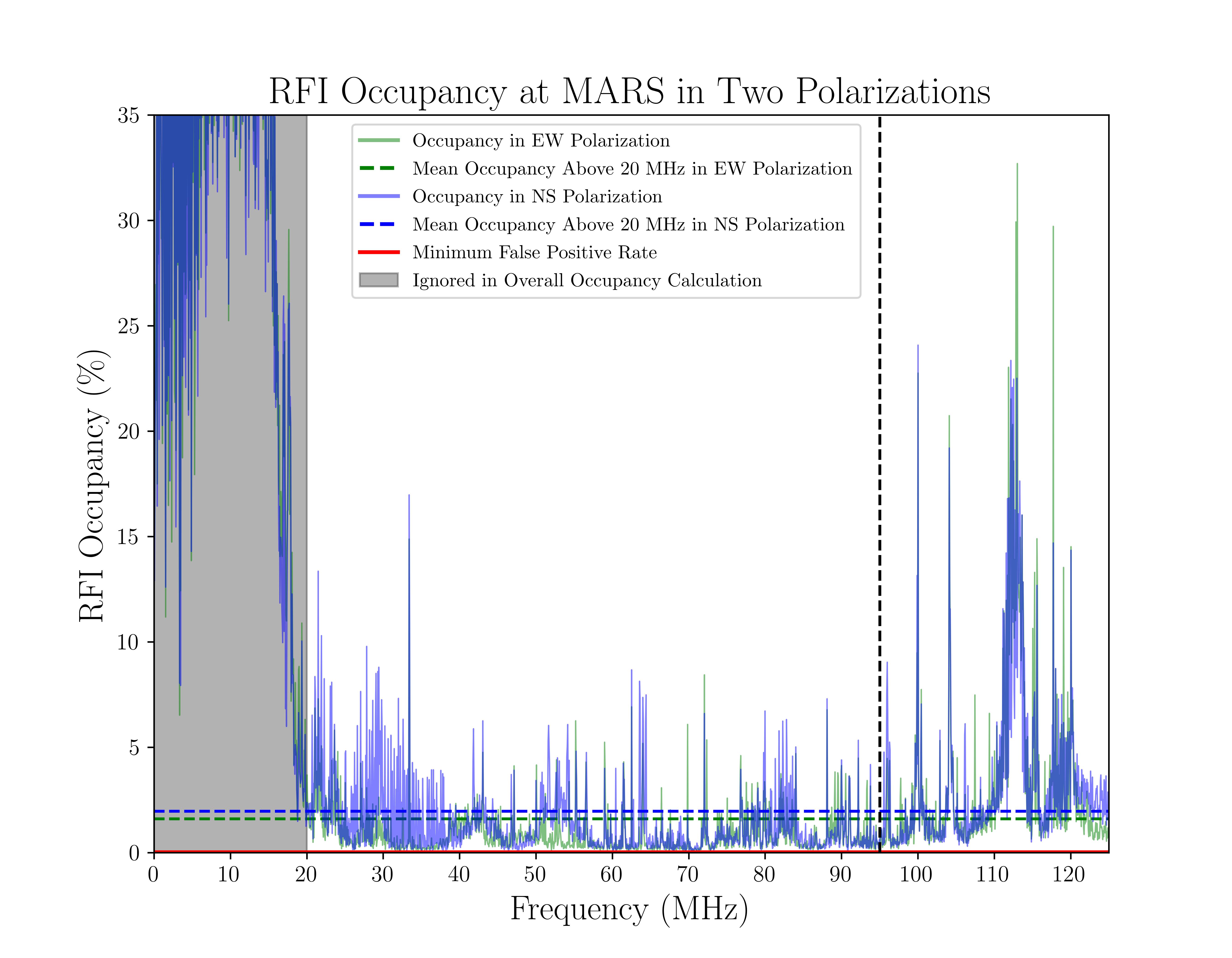}
\end{center}
\caption{Overall RFI occupancy at each frequency channel from data taken at the \mars. The green and blue lines correspond to the RFI occupancy in east-west and north-south polarized data, respectively. The horizontal blue and green dashed lines show the mean occupancy across all channels above 20~MHz for each polarization. The red line indicates the $0.04\%$ minimum false flag rate expected from Gaussian noise. The dashed line at 95~MHz indicates that any signal at higher frequency could have been aliased from 125--155~MHz (see \S\ref{sec:instrument}). The gray zone indicates the frequency range not included in the overall occupancy calculations quoted in this paper.}
\label{fig:occ}
\end{figure}

\section{Results and conclusion}
\label{sec:results}

Figure~\ref{fig:occ} shows the RFI occupancy as a function of
frequency for all data recorded at the \mars\ base.  Excluding
frequencies below 20~MHz, the overall RFI occupancy is 1.6\% in 
east-west polarized data and 2.0\% in north-south polarized data.

At frequencies above the 20~MHz ionospheric cutoff, the RFI
environment of \mars\ is exceptionally clean. The RFI occupancy over the FM band (88--108~MHz),
averaged over each polarization, is 1.6\%.  Because the low-pass filter
in the readout electronics has a 155~MHz cutoff that lies above the
Nyquist frequency, signals that appear above 95~MHz may have been aliased
from 125--155~MHz. For this reason, the RFI occupancy in the FM band is an upper bound.
The most prominent RFI feature is located at
112--113~MHz and is consistent with aliased transmission from the
ORBCOMM satellite constellation, which broadcasts at 137--138~MHz.
Since higher frequency RFI is more attenuated by distance and the Galactic signal is relatively weak above 125~MHz,
we expect that the inclusion of the ORBCOMM signal is the only significant consequence of the aliasing.
Other intermittent RFI features at $\gtrsim100$~MHz are likely
air-to-ground sources such as aeronautical radionavigation or weather
and mobile satellites\footnote{\url{https://www.ic.gc.ca/eic/site/smt-gst.nsf/eng/sf10759.html}\label{ftn:alloc}}.
The full \mars\ data set exhibits a single meteor event, when distant
radio transmitting sources reflect off of the meteor's ionization trails
and appear briefly in the observed spectra~\citep{Mallama_1999}.
During the observed meteor event, which encompassed only one time
sample, a dense cluster of RFI lines appeared in the range of
70--100~MHz, corresponding to frequencies allocated to ground-based
broadcasts.

The RFI occupancy at the remote sites surrounding
\mars\ (Figure~\ref{fig:map}) is qualitatively consistent with the
measurements recorded at the base, taking into account greater uncertainty from the
significantly smaller data volumes at the remote sites and the
different response of the antenna used during surveying.  The results
suggest that locally-generated RFI at \mars\ is not visible at few-km
distances (measurements were recorded while the \mars\ power system
was turned on) and that there are no additional major local sources of
RFI.

The results presented here indicate that \mars\ is exceptionally clear
of terrestrial RFI, even during summer conditions when the RFI levels
are expected to be at their maximum.  Observations that take place
throughout the Arctic winter will further benefit from improved
ionospheric conditions and reduced levels of local RFI from the
absence of human activity at the research base.  The quiet RFI
conditions, in combination with well developed infrastructure and
regular summer access, make \mars\ an excellent candidate site for the
development of radio astronomy experiments.  \mars\ is well suited for
the deployment of antennas that occupy relatively small footprints,
and the location is therefore particularly promising for future
observations of cosmic dawn, exploratory measurements for the cosmic
dark ages, and radio transients from ultra-high energy neutrinos.

\section*{Acknowledgments}

We gratefully acknowledge the support of the Natural Sciences and
Engineering Research Council of Canada (grant numbers
RGPIN-2019-04506, RGPNS-2019-534549, funding reference number 508480)
and the Fonds de recherche du Qu\'ebec -- Nature et technologies.  We
also acknowledge the Polar Continental Shelf Program for providing
funding and logistical support for our research program, and we extend
our sincere gratitude to the Resolute staff for their generous
assistance and bottomless cookie jars.  This research was undertaken, in part, thanks to funding
from the Canada 150 Program.  This research was enabled in part by
support provided by SciNet (\url{www.scinethpc.ca}), Compute Canada
(\url{www.computecanada.ca}), and the Hippo cluster at the University
of KwaZulu--Natal.  The authors would like to thank Chris Omelon,
Wayne Pollard, and all of the \mars\ researchers for their invaluable
advice and field help.

\bibliographystyle{apj}
\bibliography{mars2019rfi}

\begin{thebibliography}{}
\expandafter\ifx\csname natexlab\endcsname\relax\def\natexlab#1{#1}\fi

\bibitem[{{Bilitza}(2018)}]{2018AdRS...16....1B}
{Bilitza}, D. 2018, Advances in Radio Science, 16, 1

\bibitem[{{Bowman} {et~al.}(2018){Bowman}, {Rogers}, {Monsalve}, {Mozdzen}, \&
  {Mahesh}}]{2018Natur.555...67B}
{Bowman}, J.~D., {Rogers}, A. E.~E., {Monsalve}, R.~A., {Mozdzen}, T.~J., \&
  {Mahesh}, N. 2018, \nat, 555, 67

\bibitem[{{Chiang} {et~al.}(2020){Chiang}, {Dyson}, {Egan}, {Eyono}, {Ghazi},
  {Hickish}, {Jauregui-Garcia}, {Manukha}, {Moso}, {Peterson}, {Philip},
  {Sievers}, \& {Tartakovsky}}]{2020arXiv200812208C}
{Chiang}, H.~C., {Dyson}, T., {Egan}, E., {et~al.} 2020, arXiv e-prints,
  arXiv:2008.12208

\bibitem[{{DiLullo} {et~al.}(2020){DiLullo}, {Taylor}, \&
  {Dowell}}]{2020arXiv200510669D}
{DiLullo}, C., {Taylor}, G.~B., \& {Dowell}, J. 2020, arXiv e-prints,
  arXiv:2005.10669

\bibitem[{{Ellingson} \& {Kramer}(2005)}]{1552313}
{Ellingson}, S.~W., \& {Kramer}, T.~C. 2005, in 2005 IEEE Antennas and
  Propagation Society International Symposium, Vol.~3A, 561--564 vol. 3A

\bibitem[{{Hickish} {et~al.}(2016){Hickish}, {Abdurashidova}, {Ali}, {Buch},
  {Chaudhari}, {Chen}, {Dexter}, {Domagalski}, {Ford}, {Foster}, {George},
  {Greenberg}, {Greenhill}, {Isaacson}, {Jiang}, {Jones}, {Kapp}, {Kriel},
  {Lacasse}, {Lutomirski}, {MacMahon}, {Manley}, {Martens}, {McCullough},
  {Muley}, {New}, {Parsons}, {Price}, {Primiani}, {Ray}, {Siemion}, {van
  Tonder}, {Vertatschitsch}, {Wagner}, {Weintroub}, \&
  {Werthimer}}]{2016JAI.....541001H}
{Hickish}, J., {Abdurashidova}, Z., {Ali}, Z., {et~al.} 2016, Journal of
  Astronomical Instrumentation, 5, 1641001

\bibitem[{{Hicks} {et~al.}(2012){Hicks}, {Paravastu-Dalal}, {Stewart},
  {Erickson}, {Ray}, {Kassim}, {Burns}, {Clarke}, {Schmitt}, {Craig},
  {Hartman}, \& {Weiler}}]{2012PASP..124.1090H}
{Hicks}, B.~C., {Paravastu-Dalal}, N., {Stewart}, K.~P., {et~al.} 2012, \pasp,
  124, 1090

\bibitem[{Mallama \& Espenak(1999)}]{Mallama_1999}
Mallama, A., \& Espenak, F. 1999, Publications of the Astronomical Society of
  the Pacific, 111, 359

\bibitem[{{Nhan} {et~al.}(2019){Nhan}, {Bordenave}, {Bradley}, {Burns},
  {Tauscher}, {Rapetti}, \& {Klima}}]{2019ApJ...883..126N}
{Nhan}, B.~D., {Bordenave}, D.~D., {Bradley}, R.~F., {et~al.} 2019, \apj, 883,
  126

\bibitem[{{Philip} {et~al.}(2019){Philip}, {Abdurashidova}, {Chiang}, {Ghazi},
  {Gumba}, {Heilgendorff}, {J{\'a}uregui-Garc{\'\i}a}, {Malepe}, {Nunhokee},
  {Peterson}, {Sievers}, {Simes}, \& {Spann}}]{2019JAI.....850004P}
{Philip}, L., {Abdurashidova}, Z., {Chiang}, H.~C., {et~al.} 2019, Journal of
  Astronomical Instrumentation, 8, 1950004

\bibitem[{{Singh} {et~al.}(2018){Singh}, {Subrahmanyan}, {Shankar}, {Rao},
  {Girish}, {Raghunathan}, {Somashekar}, \& {Srivani}}]{2018ExA....45..269S}
{Singh}, S., {Subrahmanyan}, R., {Shankar}, N.~U., {et~al.} 2018, Experimental
  Astronomy, 45, 269

\end{thebibliography}

\end{document}